\documentclass{article} 
\usepackage{iclr2020_conference,times}

\usepackage[utf8]{inputenc} 
\usepackage[T1]{fontenc}    
\usepackage{hyperref}       
\usepackage{url}            
\usepackage{booktabs}       
\usepackage{amsfonts}       
\usepackage{nicefrac}       
\usepackage{microtype}      
\usepackage[pdftex]{graphicx}
\setcitestyle{numbers}
\setcitestyle{square}
\setcitestyle{comma}
\usepackage{amsmath}
\usepackage{caption}
\usepackage{subcaption}

\newcommand{\etal}{\textit{et al. }}
\makeatletter
\newcommand{\printfnsymbol}[1]{%
  \textsuperscript{\@fnsymbol{#1}}%
}
\makeatother

\title{Modeling Cloud Reflectance Fields using Conditional Generative Adversarial Networks}

\author{Victor Schmidt\thanks{Equal contribution}\ \ , Mustafa Alghali\printfnsymbol{1}, Kris Sankaran \& Yoshua Bengio \\
Mila, Université de Montréal\\
\texttt{\{schmidtv, sankarak\}@mila.quebec, mustafa.alghali@umontreal.ca} \\
\And
Tianle Yuan \\
NASA Goddard Space Flight Center \\
University of Maryland Baltimore County\\
\texttt{tianle.yuan@nasa.gov} \\
}

\iclrfinalcopy 
\begin{document}

\maketitle

\begin{abstract}
    We introduce a conditional Generative Adversarial Network (cGAN) approach to generate
    cloud reflectance fields (CRFs) conditioned on large scale meteorological variables
    such as sea surface temperature and relative humidity. We show that our trained model can generate
    realistic CRFs from the corresponding meteorological observations, which represents a step 
    towards a data-driven framework for stochastic
    cloud parameterization.
\end{abstract}


\section{Introduction}

Global Climate Models (GCMs) are one of the most important tools available to understand and anticipate the consequences of climate change, including changes in precipitation, increases in temperatures, and acceleration in glacial melting~\cite{climate_models_good}. One of the key physical principles these models rely on is the Earth's energy balance~\cite{north1981energy}: in short, the difference between how much energy the earth receives and how much it emits. In this context, it is paramount to model clouds accurately as they both reflect energy coming to the Earth and the infrared radiations it radiates~\cite{ramanathan1989cloud}. However, as physical processes at play in cloud composition and evolution typically range from $10^{-6}$ to $10^6$m, direct simulation of their behavior can consume up to 20\% of a GCM's computations - depending on their time and spatial scales~\cite{c_models_limitations_1,c_models_limitations_2,c_models_limitations_3}. Various efforts have tried to address this challenge. This includes traditional approaches that incorporate domain knowledge to build and validate model hypotheses using observations as well as sub-grid cloud modeling (known as super-parameterization). Alternatively, recent machine learning approaches use meteorological variables to model sub-grid clouds, thereby reducing the computational cost of super-parameterization \cite{ML_for_clouds_1, ML_for_clouds_2, ML_for_clouds_3, aicd2}.

In this paper, we extend \cite{aicd2}, a data-driven approach to contribute to cloud modeling, focusing on one of the main features used in energy balance calculations: reflectance fields. We use Conditional Generative Adversarial Networks~\cite{isola2017image} to generate these reflectance fields conditioned on meteorological variables\footnote{The code is available on Github: \url{https://github.com/krisrs1128/clouds_dist}}. We suggest using these generated images to extract important cloud parameters such as optical depth.
We believe our approach is a step towards building a data-driven framework that can reduce the computational complexity in traditional cloud modeling techniques.

Our goal is to model reflectance fields, which in turn could be used as a proxy for cloud optical depth, a major component of GCMs' energy balance computations~\cite{hartmann1992effect, corti2005mean}. To do so, we leverage 3100 aligned sample pairs $\mathcal{X} = \{r_i, m_i\}$, where  meteorological data $m_i$ are collocated with reflectances $r_i$. Each $m_i$ is a $44\times256\times256$ matrix, representing 42 measurements from MERRA-2~\cite{merra2} (see Table \ref{data-description-table}) along with longitude and latitude to account for the Earth's movement relative to the satellite\footnote{As the earth rotates, the actual geographical locations on Earth change pixel position in the data.}. On the other hand $r_i$, is a $3\times256\times256$ matrix representing each location's reflectance at RGB wavelengths (680, 550 and 450 $nm$) as measured by the Aqua dataset~\cite{aqua}. One could consider working in a Supervised Learning setting to learn a deterministic mapping $f: m_i \mapsto r_i$ ; however given the chaotic nature of climate, we need not a point estimate of the potential cloud distribution on earth, but rather an ensemble of likely scenarios given initial conditions. This motivates a generative approach using conditional GANs.

\section{Modeling reflectance fields}

\subsection{Network}

\textbf{Architecture}: motivated by Ronneberger \etal in \cite{UNet}, we use a U-Net as conditional generator. The U-Net architecture helps our generator capture global context, and skip connections allow localization. 
All of the convolution modules in our U-Net implementation consist of the same building blocks: a 3x3 convolutional layer followed by padding - which eliminates the need for cropping - followed by batch normalization, leaky ReLU, and a dropout layer with 0.25 probability. 

\textbf{Source of stochasticity}: \label{sochastic} we introduce stochasticity in the generator only through the dropout layers at both training and test times, i.e we do not use noise input vectors. As observed by Isola \etal \cite{i_to_i} dropout introduces diversity in the output of conditional GANs

\textbf{Checkerboard artifacts}: a direct implementation of this generator results in checkerboard artifacts, a result of the use of transposed convolutions to upsample the feature maps in the U-Net expansion path \cite{artifacts}. We solve this problem by replacing transposed convolution with a resize operation of the feature maps using 2d nearest neighbor interpolation, followed by a convolution as proposed in \cite{artifacts}.

\textbf{Discriminator}: we use a multi-scale discriminator as proposed by Wang \etal in \cite{multi-discriminator}. This introduces 3 discriminators with identical network structure operating on different input scales: one discriminator operates on the raw input image, while the other two operate on the raw image downsampled by factors of 2 and 4, using average pooling with a stride of 2. The motivation behind using discriminators at different scales is to provide the generator with better guidance both in the scale of global context and finer details in the image.

\subsection{Training}

\paragraph{Training objectives}
To train our generator, we use a weighted objective function composed of two losses: a non-saturating GAN loss and a matching loss:

\begin{enumerate}
    \item \emph{Non-saturating adversarial loss}.
We experimented with two types of adversarial loss: the hinge loss of \cite{Geometric-GAN} and the least square loss (LSGAN) of \cite{LS-loss}. We observe better performance with LSGAN. 
In figure \ref{fig:LS vs hing}, we also see that least squares loss is more stable during training.
    
    \item \emph{$L^1$ matching loss}. We use $L^1$ loss between generated and true reflectance images, which encourages the generator to produce outputs close to the observed images from a regression perspective. The $L^1$ loss has been found to produce less blurry outputs than $L^2$ loss \cite{i_to_i}.
\end{enumerate}

\paragraph{Optimizer}.
As we explored various optimization strategies and regularization methods, we observed significant improvement both in terms of convergence and in the quality of the generated output by using the Extra-Adam\footnote{See code at \href{https://github.com/GauthierGidel/Variational-Inequality-GAN/blob/master/optim/extragradient.py}{https://github.com/GauthierGidel/Variational-Inequality-GAN}} method proposed \cite{extra-adam} compared to Adam and SGD, see Figure~\ref{fig:extra-gradient}.




\section{Results and Discussion}
\subsection{Visual analysis}
\label{sec:visual}
\begin{figure}
    \centering
     \begin{subfigure}[b]{0.432\textwidth}
         \centering
         \includegraphics[width=\textwidth]{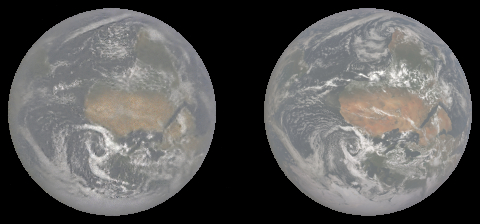}
         \caption{}
         \label{fig:results s1}
     \end{subfigure}
     \begin{subfigure}[b]{0.45\textwidth}
         \centering
         \includegraphics[width=\textwidth]{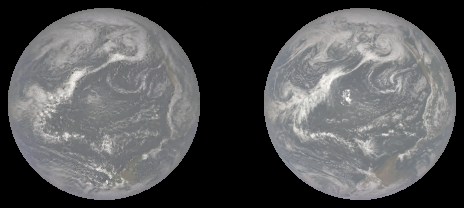}
         \caption{}
         \label{fig:results s2}
     \end{subfigure}
     \begin{subfigure}[b]{0.45\textwidth}
         \centering
         \includegraphics[width=\textwidth]{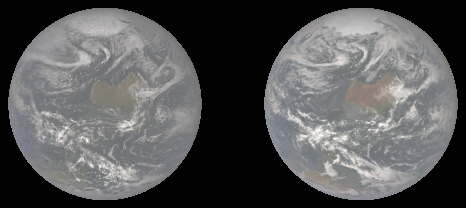}
         \caption{}
         \label{fig:results s3}
     \end{subfigure}
     \begin{subfigure}[b]{0.43\textwidth}
         \centering
         \includegraphics[width=\textwidth]{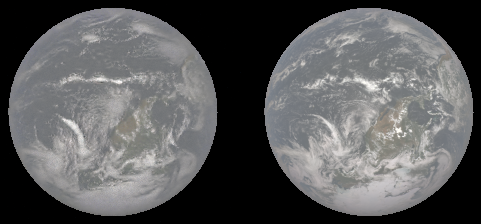}
         \caption{}
         \label{fig:results s4}
     \end{subfigure}
    \caption{4 inferences obtained from our trained model. Generated images are on the left and the corresponding true images on the right. We can see how composition is preserved, most large cloud fields have similar shapes but differ in the details.}
\label{fig:4inferences}
\end{figure}

Our U-Net generator, trained against a multi-scale discriminator and optimized by Extra-Adam, is able to generate visually appealing CRFs that are difficult to distinguish from true samples. On a validation set of ground truth images, we obtain an $\ell^2$ loss $\sim 0.027$. Figure \ref{fig:4inferences} shows 4 different pairs $(G(m_i), r_i)$: we can see that the model is able to pick up large-scale cloud structures as well as the continents and oceans beneath them. Although not as precise as the ground-truth $r_i$, the generated samples exhibit similar global composition as well as local structures.

\begin{figure}
\centering
     \includegraphics[width=0.8\textwidth]{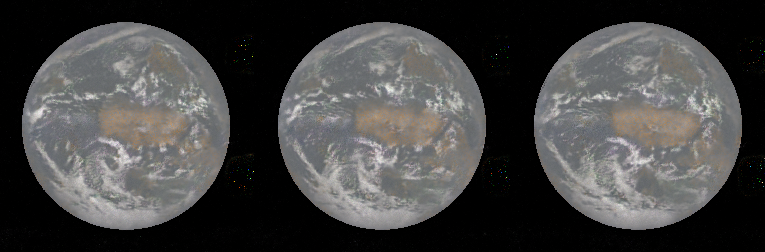}
     \caption{Reflectance fields generated by conditioning on the same input (noise comes from dropout, which is kept at test time).}
     \label{fig:results ensemble}
\end{figure}

In Figure \ref{fig:results ensemble}, we generate 3 reflectance fields from the same conditioning measurements. We notice a consistent global pattern in the three samples, with variations visible in finer details. In order to quantitatively measure diversity across generations, we fix the validation set to 5 samples that are selected manually to capture different regions of the rotating earth and generate 15 samples in total: 3 for each validation sample. For each set, we compute 3 metrics: pixel-wise mean, standard deviation and inter-quartile range across samples. Figure \ref{fig:results ensemble} shows that the model can obtain high image quality and proximity to the original distribution, but only the cost of low diversity.

Our model still has limitations, such as blurriness and small size checkerboard artifacts. We believe the reasons for this are:

\begin{enumerate}
    \item \emph{More training samples} are needed to represent such a high dimensional distribution, i.e 3100 samples are not enough to train a deep U-Net generator ($\sim 1.4$ million parameters) and discriminator ($\sim 8.3$ million parameters).

    \item  \emph{More hyperparameter tuning}, including architectural choices of the generator and discriminator to ensure the right capacity balance that lead to a long lasting game and avoid prematurely saturated learning.
    
    \item \emph{Further training} -- we can see that the discriminator loss still slightly oscillates after  saturation points and eventually decreases with number of steps as shown in \ref{fig:saturating d}.
\end{enumerate}

\begin{figure}[h!]
\centering
    \includegraphics[width=0.6\textwidth]{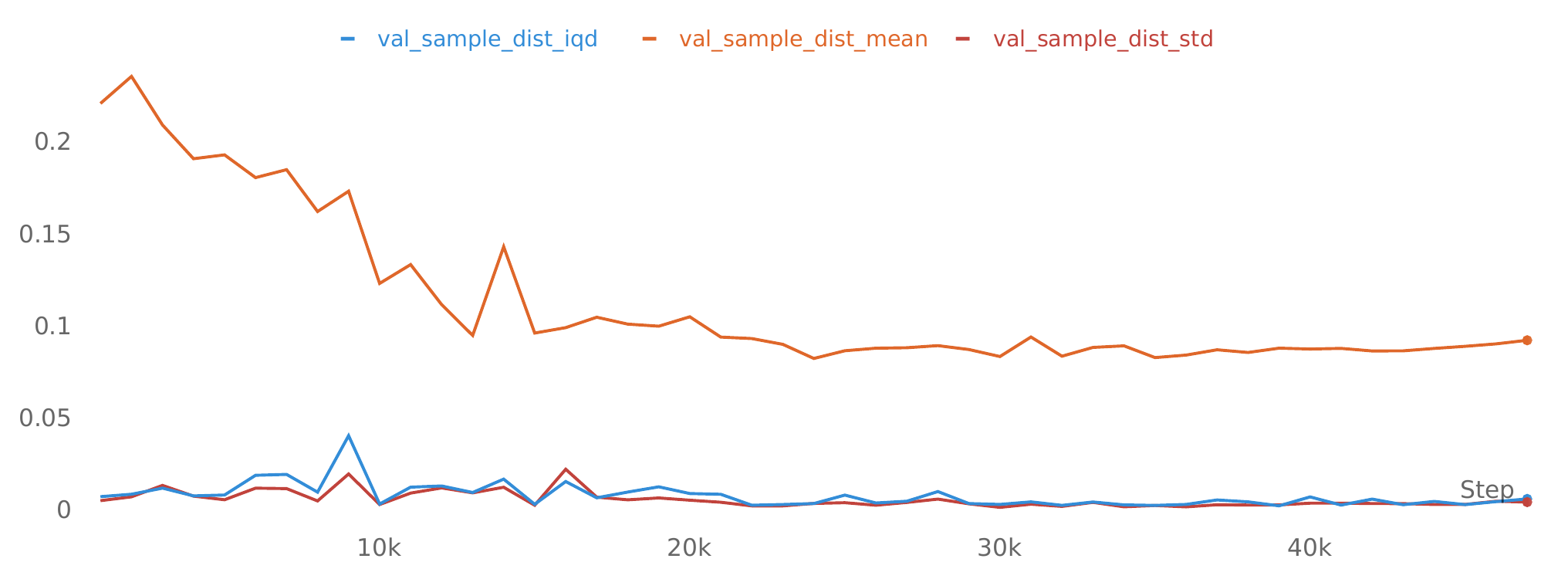}

    \label{fig:iqr}
    \caption{Plot of the inter-quartile distance, mean, and standard deviation of 15 generated samples at different steps during the model's training.}
    \label{fig:visualization_results}
\end{figure}     

\subsection{Spectral analysis}
\begin{figure}[h!]
    \centering
    \includegraphics[width=0.7\textwidth]{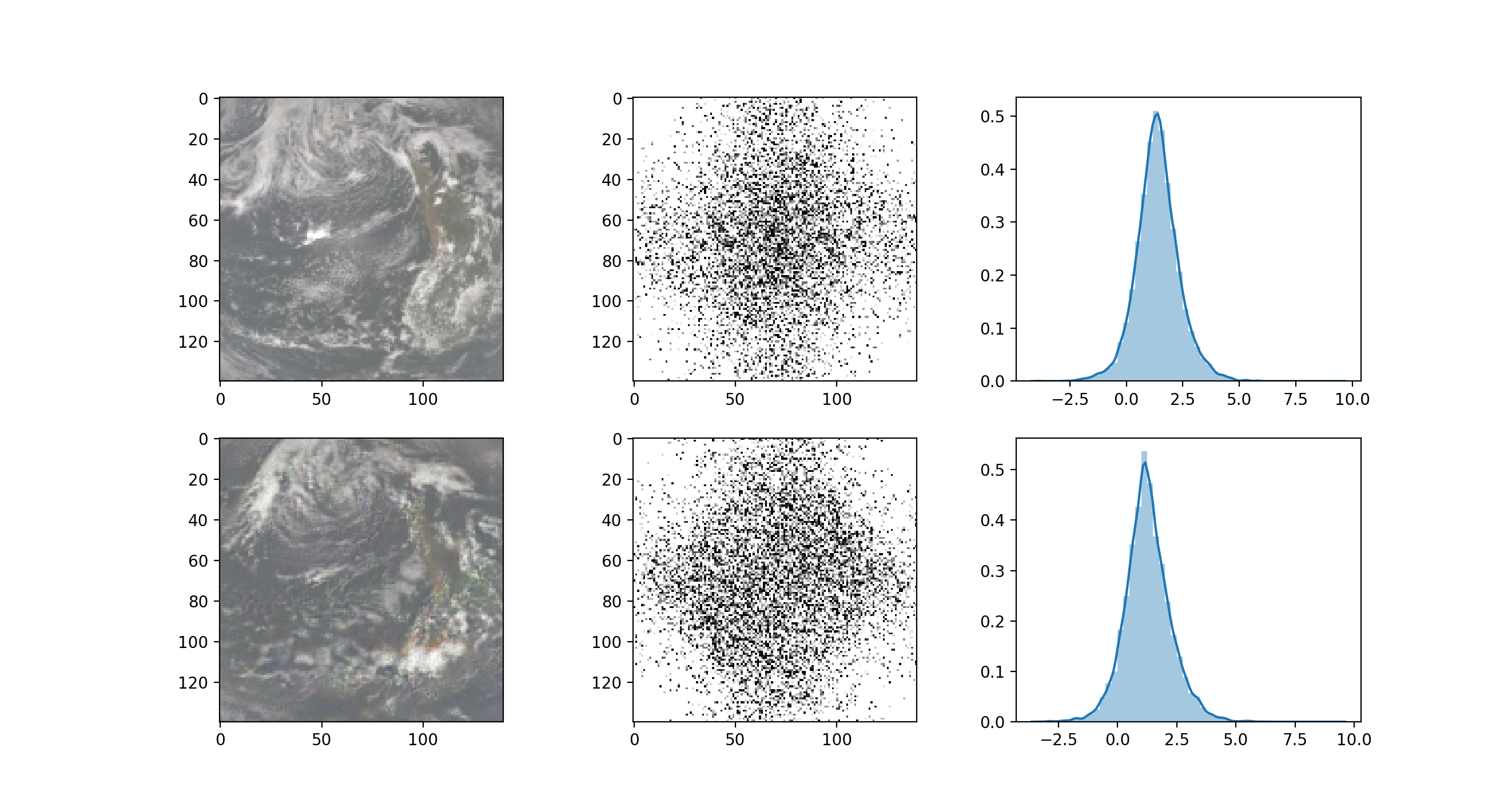}
    \caption{Comparison of DFTs (from left to right: image, frequency magnitudes, histogram of magnitudes), with real data in the top row and a generated reflectance field (from the real data's associated measurements)in the bottom one}
    \label{fig:DFT}
\end{figure}

Although visual inspection techniques can give insight into GAN performance, it is an expensive, cumbersome, and subjective measure~\cite{borji2019pros}. We address this issue by comparing the frequency spectrum of true and generated samples using 2D Discrete Fourier transform (DFT). This allows us to compare the images' geometric structures by examining the contribution of frequency components~\cite{DSP}. 
We compare the the magnitudes of the 2D DFT calculated from the grayscale versions of the true and generated images, and compare the histograms of the calculated magnitudes, their means, variances and the logarithmic average $L^2$ distances. In figure \ref{fig:DFT}
we observe that our generated images have consistent and similar DFT distributions to those of their corresponding true images, with a very small average $L^2$ distance. 

\section{Conclusion and future work}
We show that using conditional GANs to model CRFs can be an effective approach towards building a data-driven framework. 
We think our approach could significantly help improve the computation time of clouds modeling in global climate models.
\newline
Future work includes increasing the size of the dataset and exploitation of the temporal structure in our data in two ways: by adding date and time as extra labels to the input variable, and by using temporal cross validation \cite{nested-cross-validation} to validate our generator's ability to predict possible changes in cloud distribution over time. We also plan to increase the diversity in the generated ensembles by  incorporating input noise channels as an extra source of stochasticity. To address what we suspect to be mode collapsing in our network (the matching loss discourages the exploration of other potential modes in the data) we suggest using staged training where an adaptive weight for the matching loss encourages the generator to regress onto true images during early stages of the training, eventually decreasing to zero as training progresses.

\bibliography{iclr2020_conference}
\bibliographystyle{iclr2020_conference}
\clearpage
\appendix
\section{Data processing}

\begin{figure}[h!]
  \centering
  \includegraphics[width=0.5\linewidth, height = 0.4\linewidth]{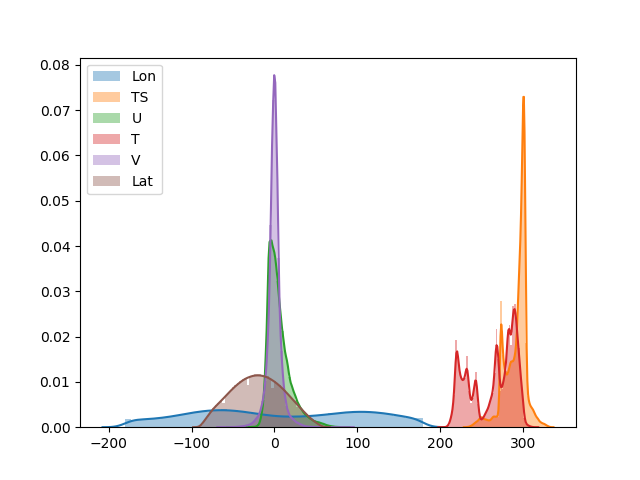}
  \caption{Histograms of six input variables shows the variance in scales.}
  \label{fig: hist}
\end{figure}

During data processing, we Winsorize reflectance data to remove artifacts that are present in some training samples due to sensor noise, clipping values to the 95th percentile for each channel. This is necessary as meteorological variables have different scales, see Figure ~\ref{fig: hist}. We standardize channels to have values in $\left[-1, 1\right]$ and zero mean. In order to to avoid introducing unnecessary bias from values outside the earth disk, we first crop the images to cut off most of these values and then upsample them again to their original size using  2D nearest neighbors, replacing the remaining values with $-3$ (mean - 3x standard deviation) to avoid introducing any unnecessary bias in the data distribution.

We used running statistics to compute the summary statistics of the data due to the huge size of input tensors, which do not fit in 16GB GPU memory at one time. We also increased the number of data loader workers to 12; this accelerates the data loading process by $6\times$.

\begin{table}[h!]
  \caption{Description of input components}
  \label{data-description-table}
  \centering
  \begin{tabular}{lll}
    \toprule
    \cmidrule(r){1-2}
    Name     & Description     & Number of channels \\
    \midrule
    U, V   & Wind components in 10 atmospheric levels  &  20    \\
    T  & Temperature in 10 atmospheric levels & 10 \\
    RH     & Relative-humidity in 10 atmospheric levels & 10      \\
    SA     & Scattering angle  & 1 \\
    TS     & Surface Temperature & 1 \\
    Lat, Long & Latitude and Longitude & 2 \\
    \bottomrule
  \end{tabular}
\end{table}

\clearpage
\section{Hyper-parameter comparisons}

\begin{figure}[h!]
     \centering
     \begin{subfigure}[b]{0.4\textwidth}
         \centering
         \includegraphics[width=\textwidth]{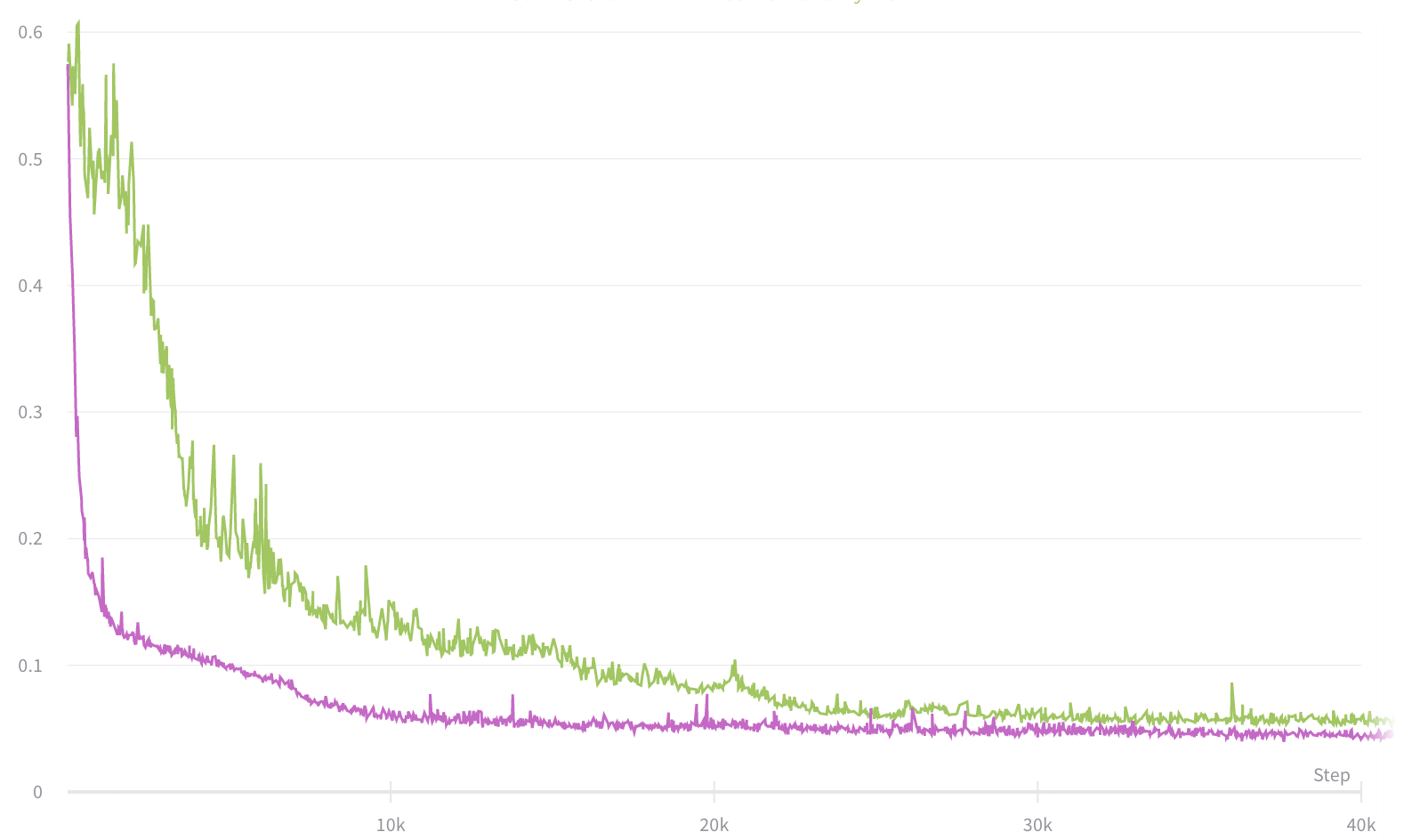}
         \caption{L1 matching loss}
         \label{fig:matching_loss_c}
     \end{subfigure}
     \begin{subfigure}[b]{0.4\textwidth}
         \centering
         \includegraphics[width=\textwidth]{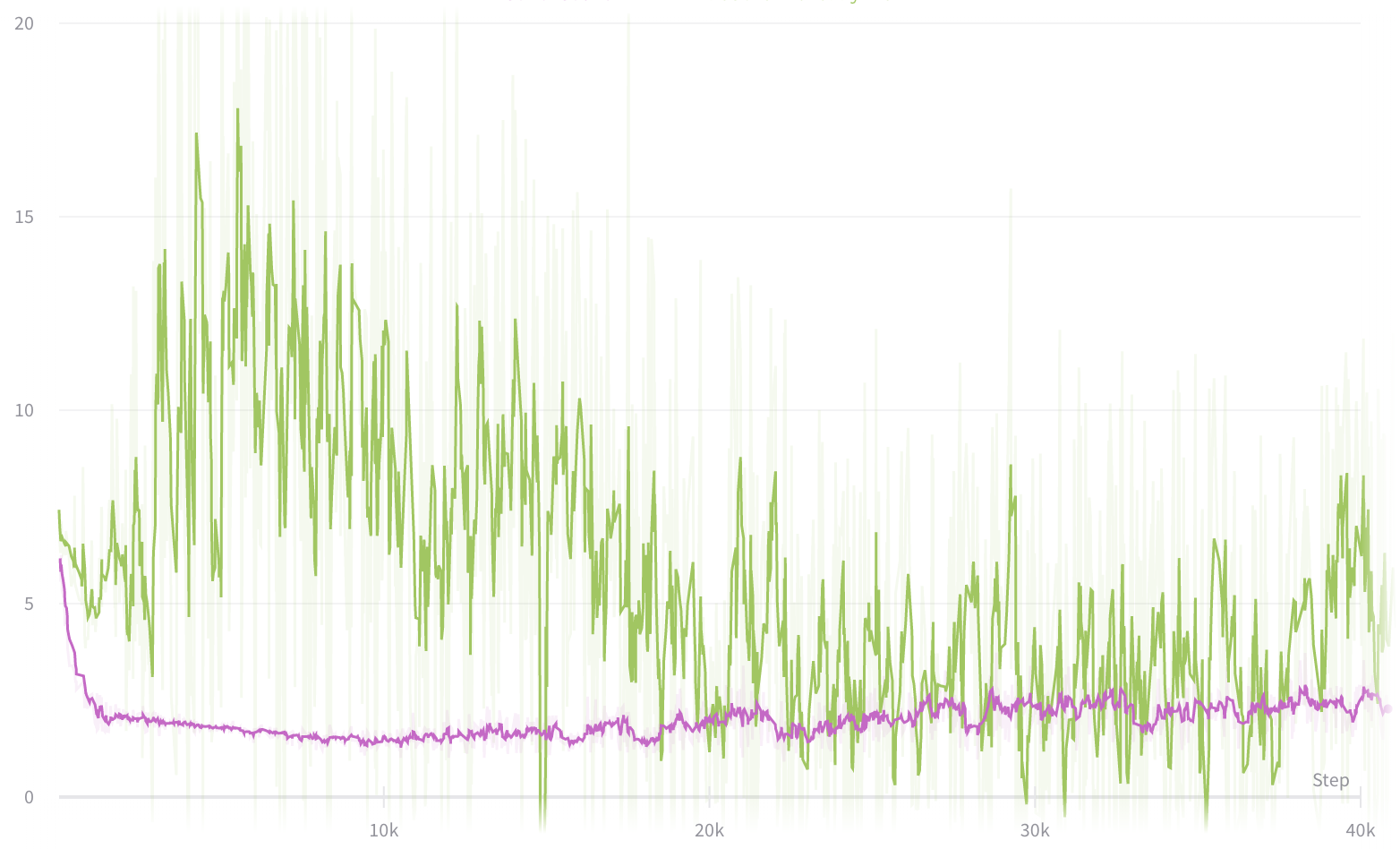}
         \caption{Total weighted generator loss.}
         \label{fig:total_g_loss_c}
     \end{subfigure}
        \caption{Comparison between the hinge loss (green) and the least squares loss (purple) on model training stability and convergence, we observe that the latter performs better both in optimization of the $L^1$ loss and the total weighted generator loss. We configured Adam and ExtraAdam to use $\beta_1$ = 0.5, and $\beta_2$ = 0.99 in all experiments.}
        \label{fig:LS vs hing}
\end{figure}

\begin{figure}[h!]
    \centering
     \begin{subfigure}[b]{0.33\textwidth}
         \centering
         \caption{$L^1$ matching loss.}
         \label{fig:L1 optimizers}
     \end{subfigure}
     \begin{subfigure}[b]{0.33\textwidth}
         \centering
         \includegraphics[width=\textwidth]{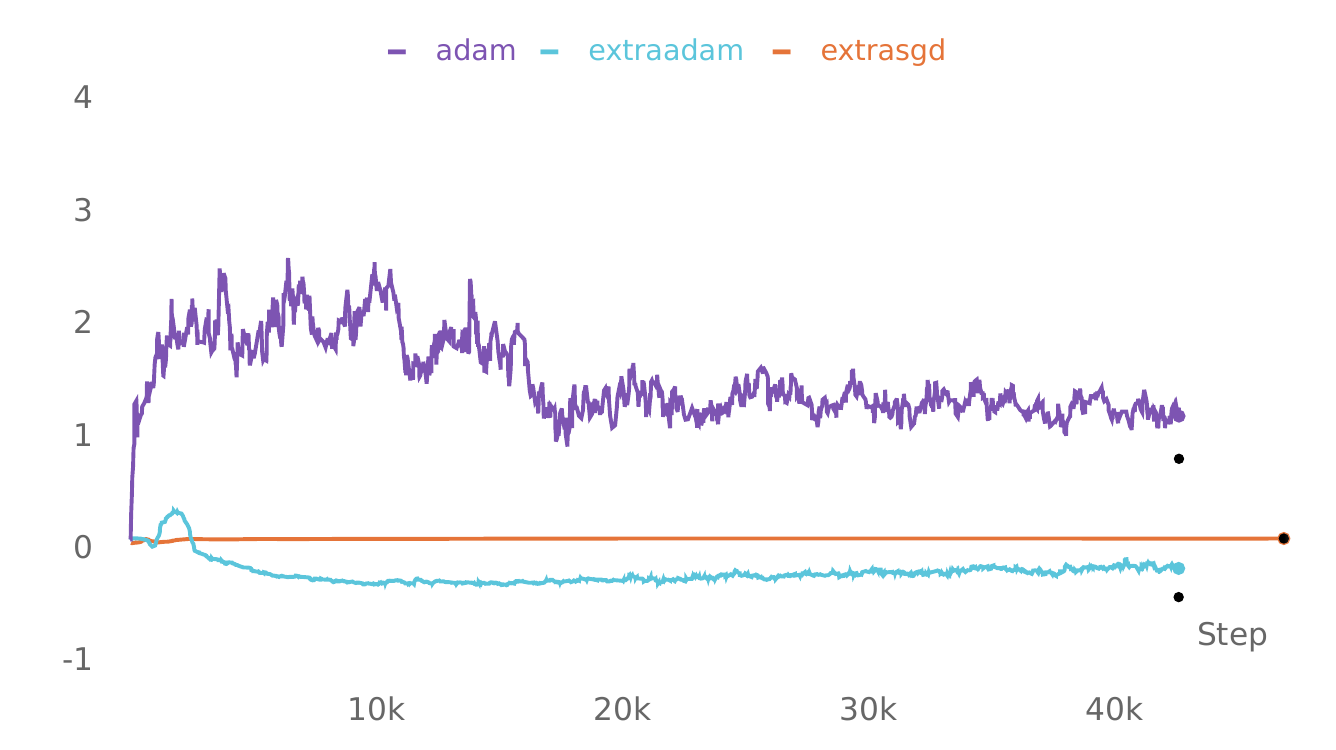}
         \caption{Generator adversarial loss.}
         \label{fig:g optimizers}
     \end{subfigure}
     \begin{subfigure}[b]{0.32\textwidth}
         \centering
         \includegraphics[width=\textwidth]{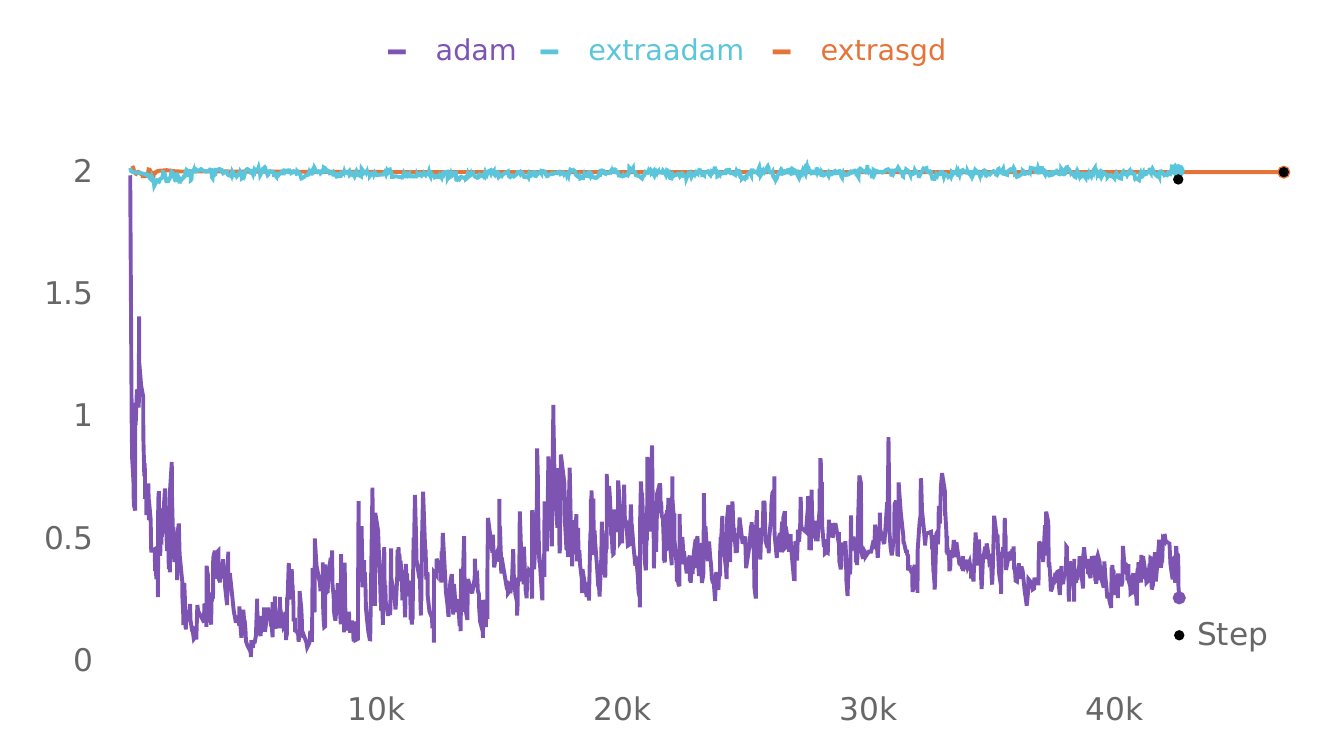}
         \caption{Discriminator loss.}
         \label{fig:d optimizers}
     \end{subfigure}
     \newline

    \begin{subfigure}[b]{0.15\textwidth}
         \centering
         \includegraphics[width=\textwidth]{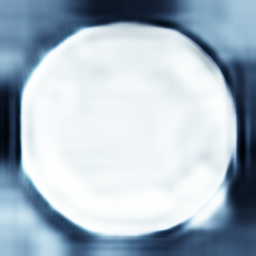}
         \caption{Adam}
         \label{fig:total_g_loss_c}
    \end{subfigure}
    \begin{subfigure}[b]{0.15\textwidth}
         \centering
         \includegraphics[width=\textwidth]{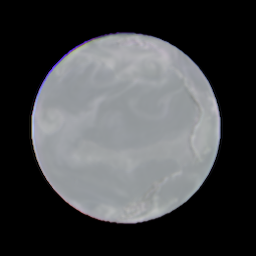}
         \caption{ExtraSGD}
         \label{fig:total_g_loss_c}
    \end{subfigure}
    \begin{subfigure}[b]{0.15\textwidth}
         \centering
         \includegraphics[width=\textwidth]{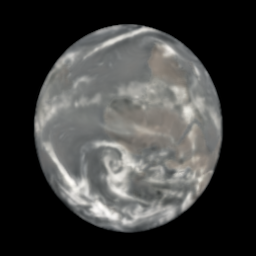}
         \caption{ExtraAdam}
         \label{fig:total_g_loss_c}
    \end{subfigure}
    \begin{subfigure}[b]{0.15\textwidth}
         \centering
         \includegraphics[width=\textwidth]{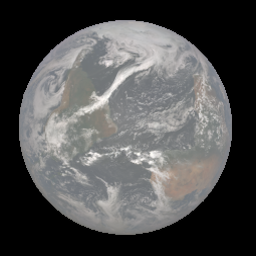}
         \caption{Real earth}
         \label{fig:total_g_loss_c}
    \end{subfigure}
        \caption{The losses of 3 experiments with 3 different optimizers: Adam, ExtraSGD, and ExtraAdam, along with generated outputs for each experiment conditioned on a fixed meteorological input. ExtraAdam shows better convergence, less oscillating losses, and more visually appealing output relative to Adam and ExtraSGD.}
        \label{fig:extra-gradient}
\end{figure}
\begin{figure}[h!]
    \centering
      \begin{subfigure}[b]{0.5\textwidth}
         \centering
         $\vcenter{\hbox{\includegraphics[width=\textwidth]{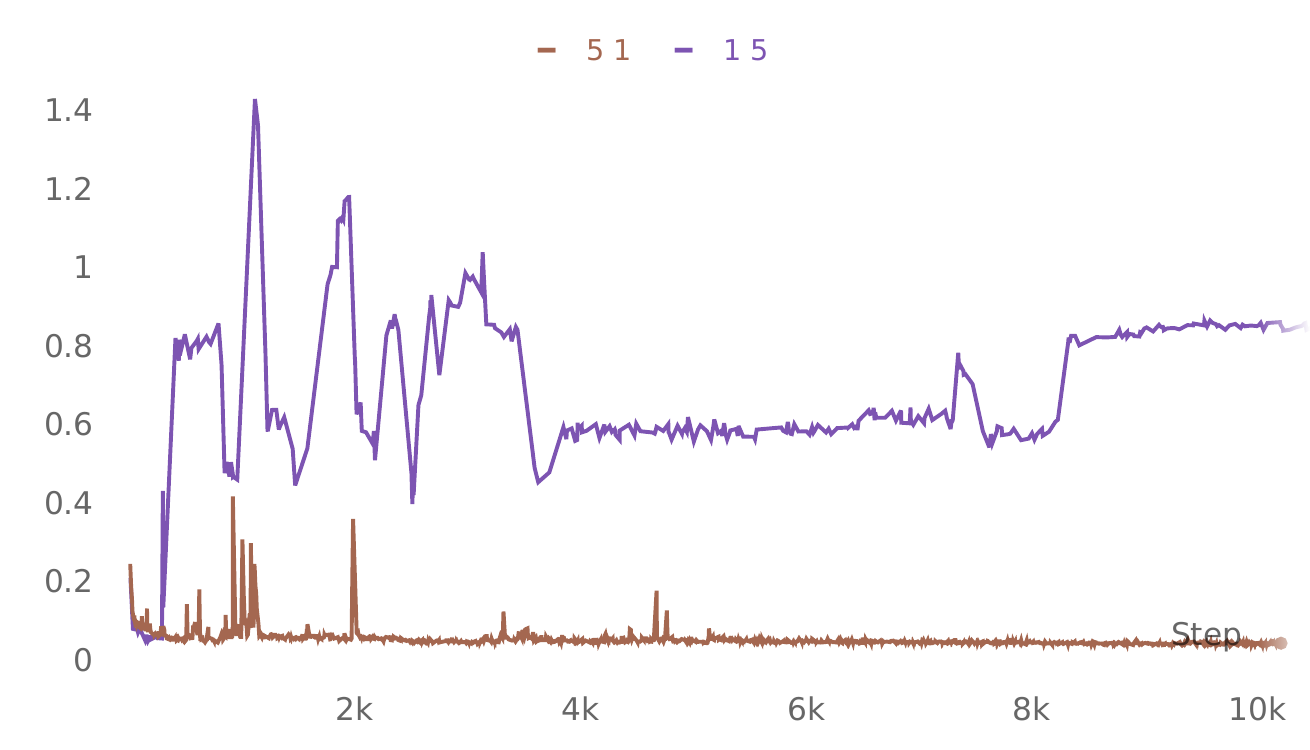}}}$
         \caption{$L^1$ matching loss}
         \label{fig:total_g_loss_c}
    \end{subfigure}
     \begin{subfigure}[b]{0.15\textwidth}
         \centering
          $\vcenter{\hbox{\includegraphics[width=\textwidth]{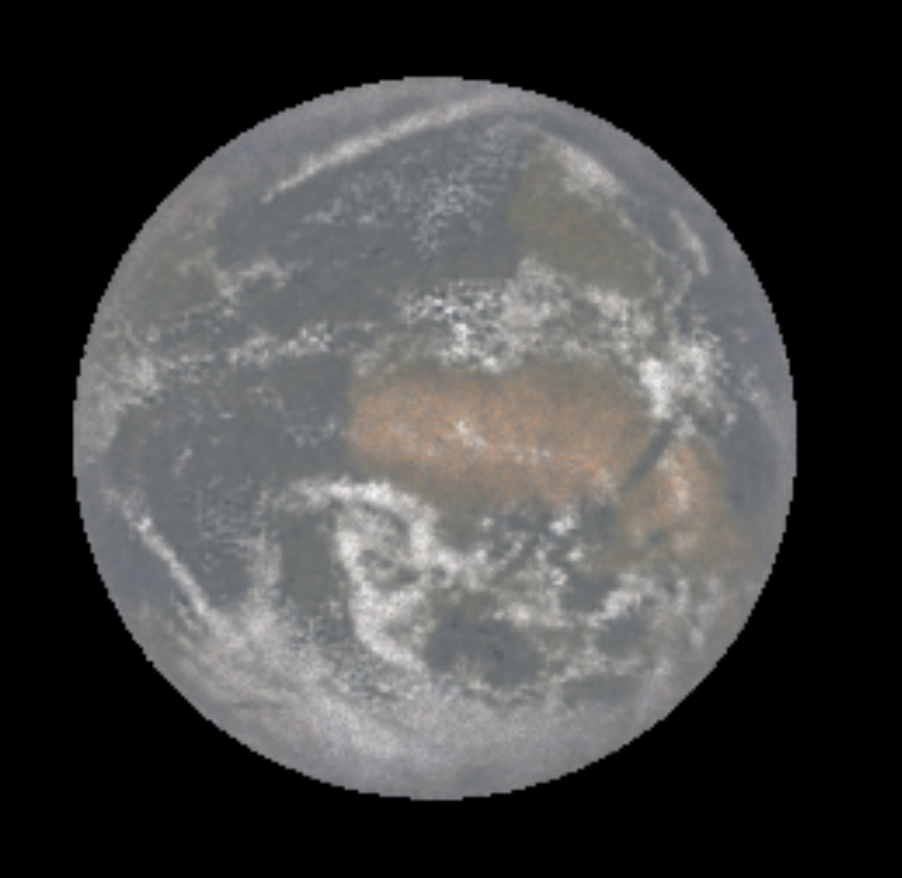}}}$
         \caption{$\lambda_1/ \lambda_2 = 5$}
         \label{fig:g optimizers}
     \end{subfigure}
      \begin{subfigure}[b]{0.15\textwidth}
         \centering
          $\vcenter{\hbox{\includegraphics[width=\textwidth]{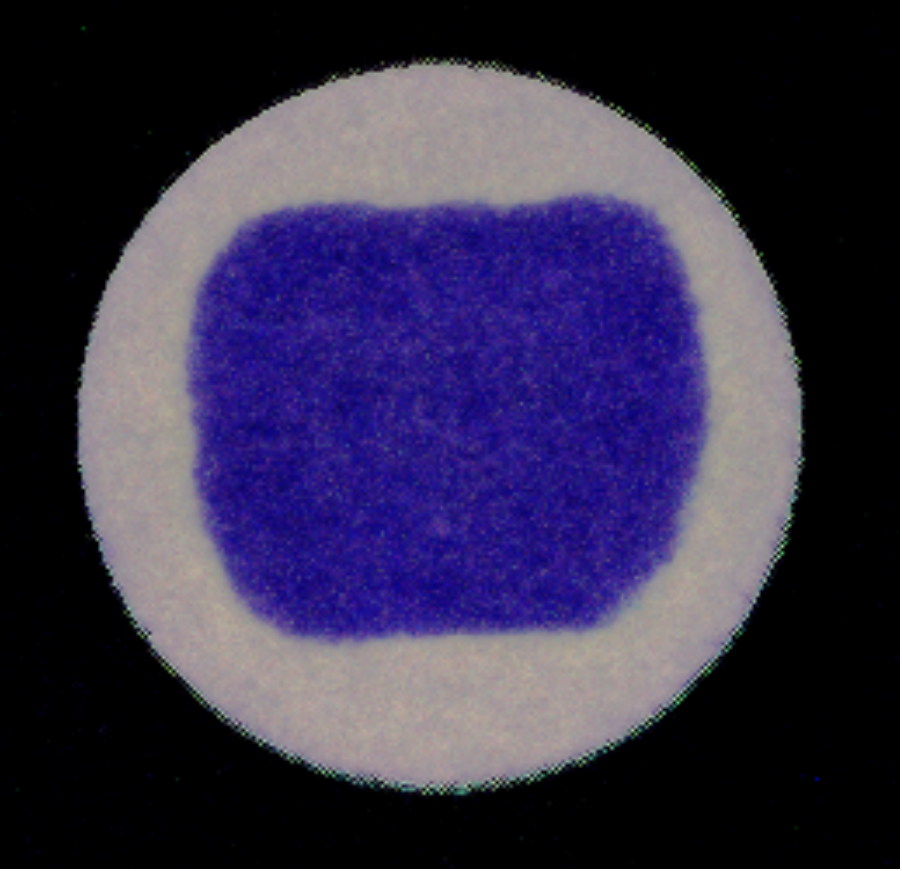}}}$
         \caption{$\lambda_1/ \lambda_2 = 0.2$}
         \label{fig:L1 optimizers}
     \end{subfigure}
     \begin{subfigure}[b]{0.1525\textwidth}
         \centering
          $\vcenter{\hbox{\includegraphics[width=\textwidth]{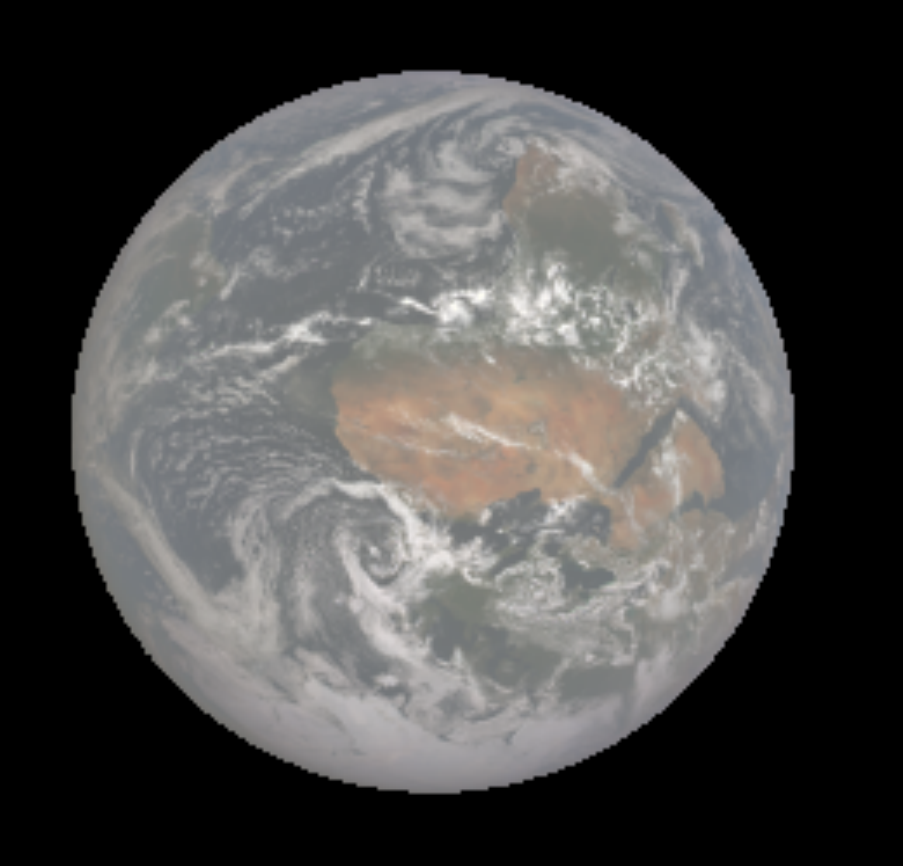}}}$
         \caption{real}
         \label{fig:d optimizers}
     \end{subfigure}
        \caption{Two experiments with two different $\lambda_1/\lambda_2$ ratios shows stability of the discriminator loss for larger ratios -- resulting in more realistic images, see (b) --, while small ratios produce hallucinated and unrealistic images, see (c).}
        \label{fig:lambdas}
        
\end{figure}

 \paragraph{Regression vs. hallucinated features}
 The $\lambda_1 / \lambda_2$ ratio in the generator's weighted objective function plays an important role in the behavior of our generator; experiments with large ratios in the range of $\left[1, 10\right]$ behave like supervised models where we regress the generated images with $L^1$ loss, while small ratios of $\leq 0.5$ tend to give the generator more freedom to explore the distribution of interest, without being penalized for not matching low frequency details. This causes the generator to hallucinate features that do not exist in the true images (Figure \ref{fig:lambdas}). This behavior matches our expectations: 3100 samples is not sufficient to learn the conditional distribution of such
 variable and high-dimensional data. 

\paragraph{Architecture Summary}

\begin{figure}
    \centering
    \includegraphics[width=0.9\textwidth]{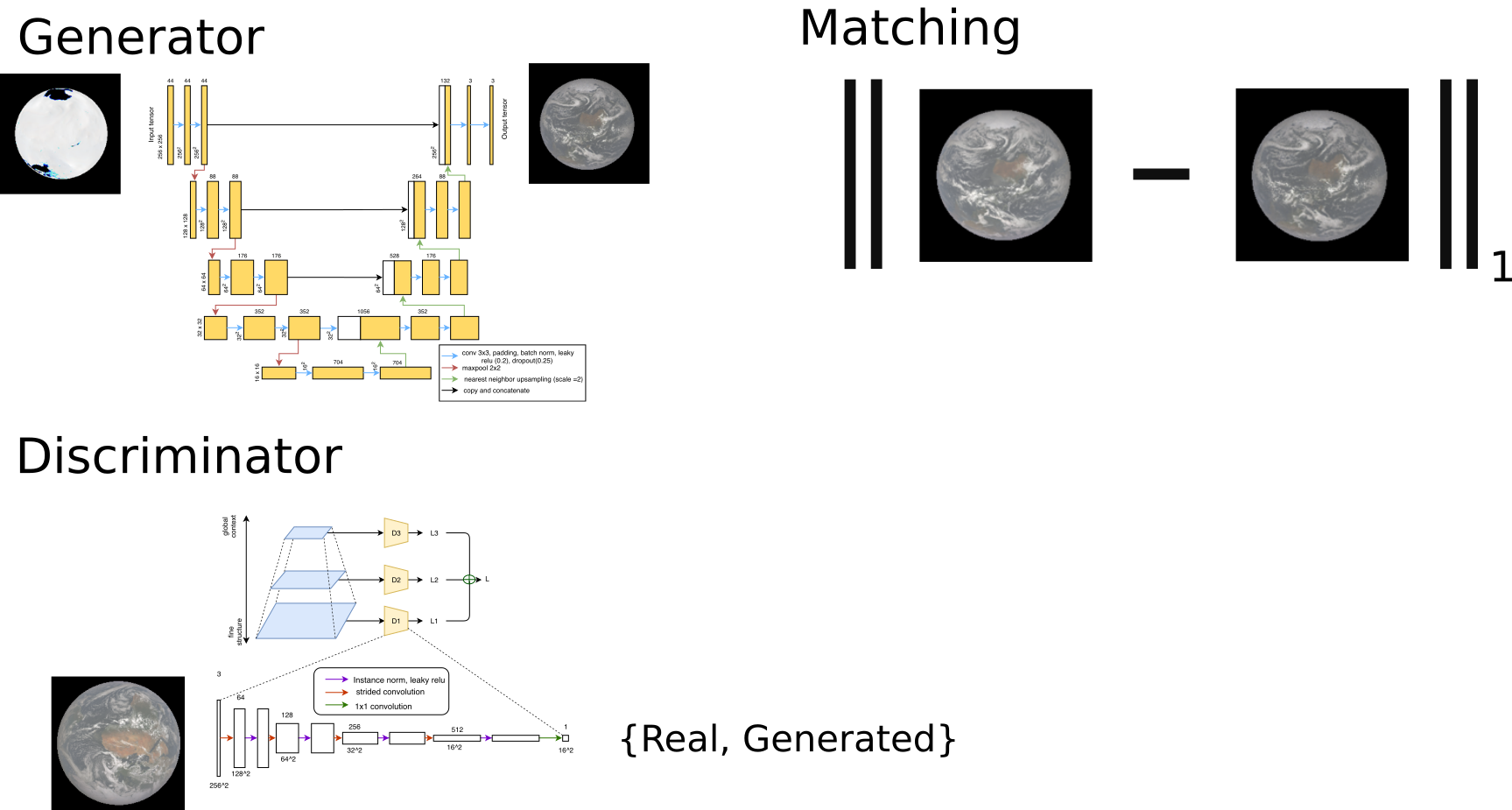}
    \caption{The overall architecture includes two models, a U-Net generator and multiscale discriminator, and the optimization objective combines a cGAN loss with an $L^1$ matching loss.}
    \label{fig:summary}
\end{figure}

The main components of our architecture are summarized in Figure \ref{fig:summary}.

\begin{figure}[h!]
    \centering
     \begin{subfigure}[b]{0.2\textwidth}
         \centering
         \includegraphics[width=\textwidth]{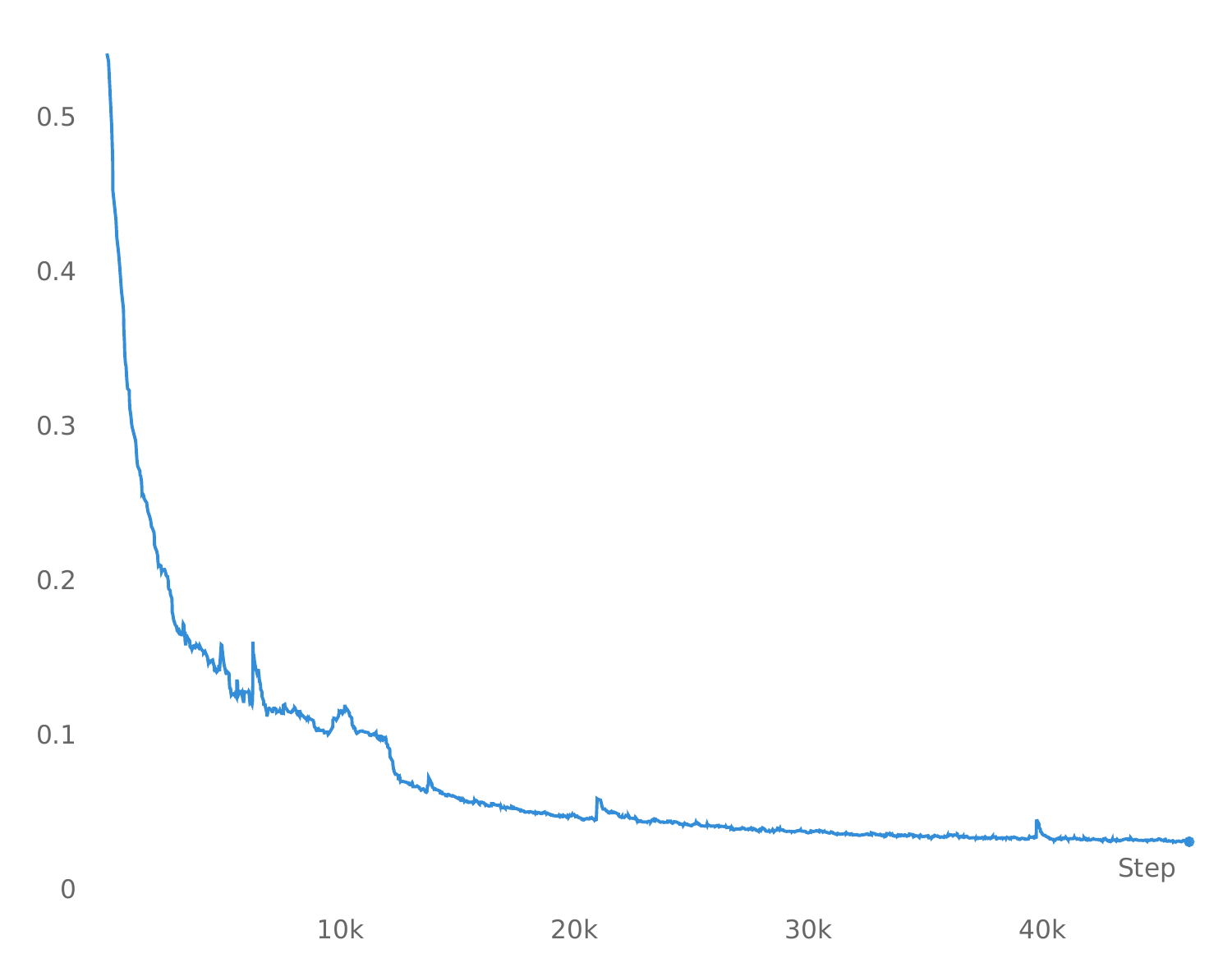}
         \caption{$L^1$ matching loss}
         \label{fig:L1 optimizers}
     \end{subfigure}
     \begin{subfigure}[b]{0.2\textwidth}
         \centering
         \includegraphics[width=\textwidth]{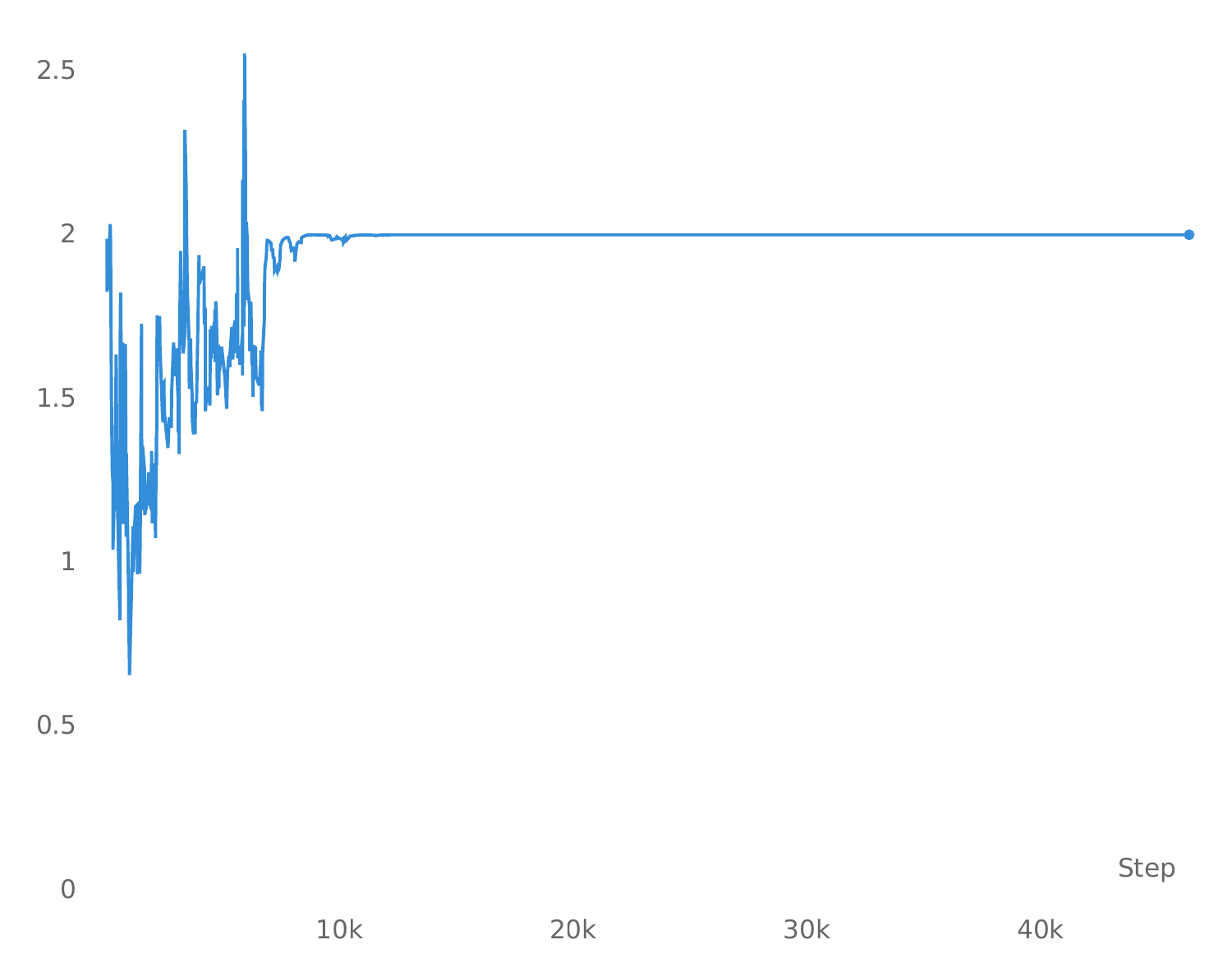}
         \caption{Generator loss}
         \label{fig:g optimizers}
     \end{subfigure}
     \begin{subfigure}[b]{0.2\textwidth}
         \centering
         \includegraphics[width=\textwidth]{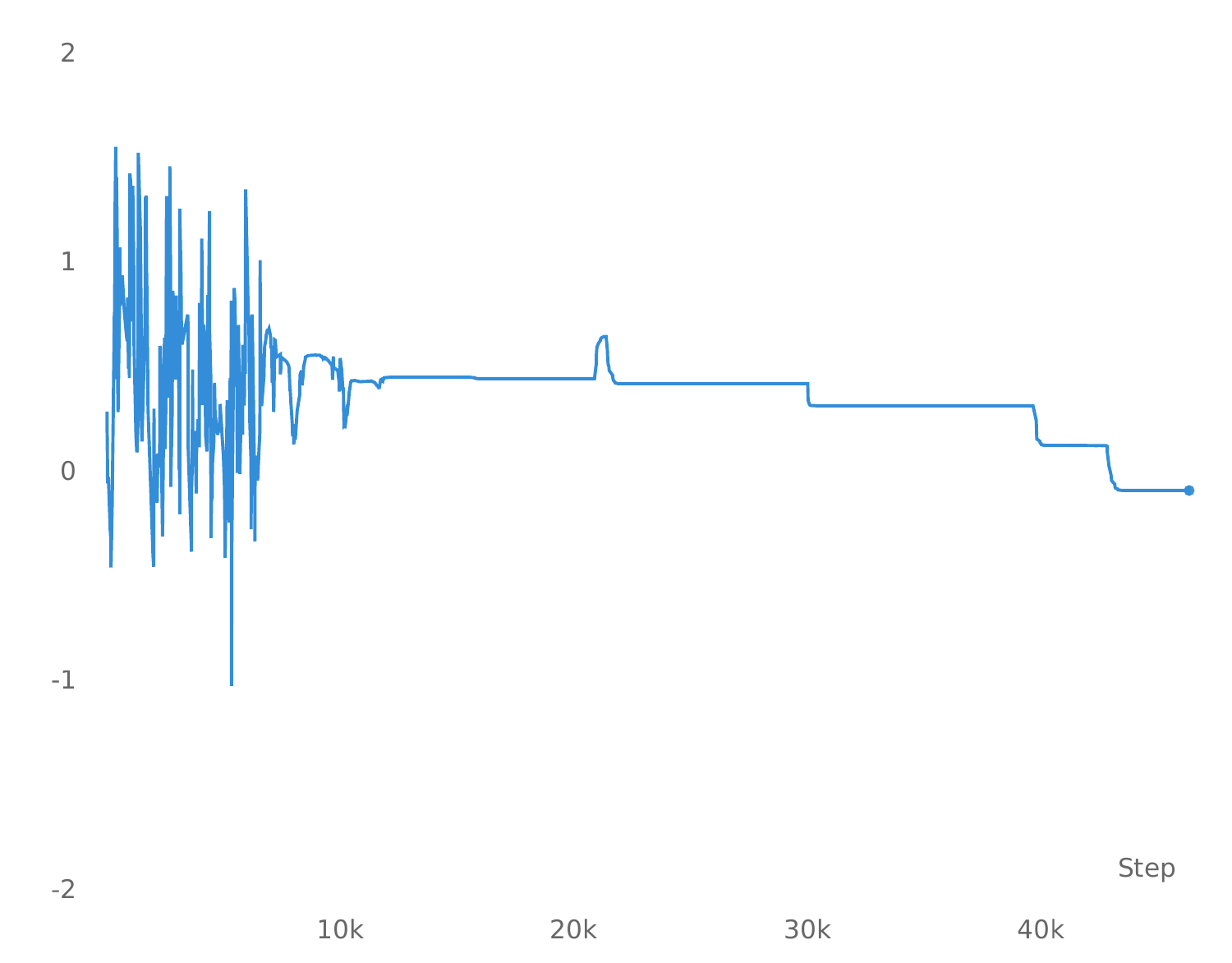}
         \caption{Discriminator loss}
         \label{fig:saturating d}
     \end{subfigure}
     \begin{subfigure}[b]{0.15\textwidth}
         \centering
         \includegraphics[width=\textwidth]{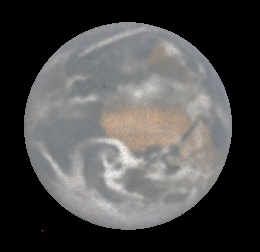}
         \caption{fake}
         \label{fig:d optimizers}
     \end{subfigure}
     \begin{subfigure}[b]{0.15\textwidth}
             \centering
             \includegraphics[width=\textwidth]{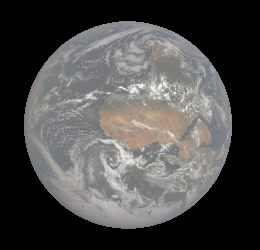}
             \caption{real}
             \label{fig:d optimizers}
    \end{subfigure}     
        \caption{An example experiment with a large discriminator learning rate (0.001) shows early saturation of the discriminator loss after $\sim$ 10K steps, resulting in an idle generator loss starting from the same point, the generated sample thus looks blurry when compared to the real one.}
        \label{fig:saturating}
        
\end{figure}

\paragraph{Sharpness of generated images}
Generating sharp images that can show complicated and detailed clouds structures such as spinning clouds is both important and challenging.
We address this challenge by carefully choosing the discriminator learning rate to avoid saddle point convergence and non-convergence, which result in bad generations. We hypothesize that the generator might have not been trained long enough to learn such micro-level details and thus generate blurry output. In most of these cases, we observe the generator is the dominant player, and the discriminator is fooled at saturation points, see the example shown in Figure \ref{fig:saturating}.

\end{document}